\documentclass[aps,pra,twocolumn,showpacs,letterpaper,superscriptaddress]{revtex4-1}

\usepackage[colorlinks=true, citecolor=blue, linkcolor=blue, urlcolor=blue]{hyperref}
\usepackage{graphicx,dcolumn,longtable,epsfig}
\usepackage[usenames]{color}
\usepackage{amssymb}
\usepackage{amsmath,mathtools}
\usepackage{bm}
\usepackage{footnote}
\usepackage{float}
\usepackage{subfigure}
\usepackage{color}       
\usepackage{ulem}
\usepackage[T1]{fontenc}

\newcommand{\be}{\begin{equation}}
\newcommand{\ee}{\end{equation}}

\def\bea{\begin{eqnarray}}
\def\eea{\end{eqnarray}}

\begin{document}

\title{Density instabilities and thermal stabilization of phase‑separated states in dipolar lattice bosons}
\author{Yaghmorassene Hebib }
\affiliation{Department of Physics, Clark University, Worcester, Massachusetts 01610, USA}
\affiliation{Department of Physical science, Butte College, Oroville, California 95965, USA}
\author{Stefano Peaquin}
\affiliation{Institute for Condensed Matter Physics and Complex Systems,
DISAT, Politecnico di Torino, I-10129, Torino, Italy}
\author{Chao Zhang}
\email{chaozhang@ahnu.edu.cn}
\affiliation{Department of Physics, Anhui Normal University, Wuhu, Anhui 241000, China}
\author{Vittorio Penna}
\affiliation{Institute for Condensed Matter Physics and Complex Systems,
DISAT, Politecnico di Torino, I-10129, Torino, Italy}
\author{Barbara Capogrosso-Sansone}
\affiliation{Institute for Condensed Matter Physics and Complex Systems,
DISAT, Politecnico di Torino, I-10129, Torino, Italy}

\begin{abstract}
Recent advances in realizing nearly degenerate dipolar gases in optical lattices have enabled the study of quantum systems with long-range anisotropic interactions. Here, we investigate hard-core dipolar bosons on a two-dimensional square lattice described by an extended Bose–Hubbard model. Using path-integral quantum Monte Carlo simulations at fixed azimuthal angle $\varphi=45^\circ$, we investigate density instabilities arising due to first order phase transitions. We start by mapping the ground-state phase diagram at half-filling as a function of dipolar interaction strength and polar angle $\theta$. For weak interactions, the system remains superfluid for all $\theta$. Above a critical interaction strength, the superfluid phase becomes unstable and gives way to checkerboard, stripe, or incompressible phases depending on $\theta$. 

For $\theta\gtrsim 62^\circ$, we find that half-filling becomes unstable and only empty, $n=0$, and fully filled, $n=1$, states are  stable. 
Interestingly, unlike recent experimental reports of a self-bound insulator at half filling, the homogeneous ground state does not support such a phase but rather exhibits a direct first-order transition between $n = 0$ and $n = 1$.
At finite temperature, thermal fluctuations shift the onset of density instabilities to larger $\theta$ and stabilize intermediate fillings in the regime where half-filling is unstable in the ground state. This leads to phase-separated states consisting of empty and fully filled regions that resemble the experimentally observed “self-bound insulator”. In a harmonic trap, similar structures also emerge from phase coexistence associated with the underlying first-order transition.

\end{abstract}

\maketitle

\section{Introduction}
The rapid progress in trapping and precise control of ultracold atoms and molecules has opened new avenues for the exploration of magnetic and electric dipolar interactions across a wide range of systems~\cite{Su2023, Griesmaier2005, Frisch2015, Gadway2016, Lu2009,Langen:2024aa,Chomaz_2023}. 
These platforms serve as versatile quantum simulators, enabling controlled studies of strongly correlated many-body physics with long-range interactions. A central advantage is the ability to manipulate both the strength and orientation of dipole moments via external fields, which has motivated extensive theoretical work, particularly in reduced dimensionality.
Owing to their long-range and anisotropic character, dipolar interactions have been predicted to stabilize a rich variety of exotic quantum phases and phenomena, including numerous crystalline and supersolid states, superfluidity of self-assembled chains, topological quantum phases, roton-like excitation spectra, etcetera~\cite{Koziol2024,PhysRevA.107.043318,Baranov2008,Sinha_supersolids, Capogrosso-Sansone:2011aa,PhysRevA.90.043604,lhfx-c4xr,PhysRevA.90.043635,PhysRevB.87.081106,PhysRevB.111.024511, PhysRevLett.105.135301,PhysRevA.97.013615,Yamaguchi2010, Brunn2014,Bhongale2013,RecatiStringari2023,Mukherjee2023,Zampronio2024,SOCdipolar2024,He2025}.

Recently, a variety of quantum solid phases have been experimentally observed using ultracold Erbium (Er) atoms~\cite{Su2023}. In that work, the extended Bose–Hubbard model was realized by loading magnetic atoms into an optical lattice. The tilt angle of the magnetic dipole moment was tuned via an external magnetic field, enabling systematic exploration of different tilt angles. By employing the quantum gas microscope in combination with an accordion lattice, the authors achieved site-resolved imaging of atoms on a lattice of about $15\times15$ sites in the central region of the harmonic trap. They reported the experimental observation of both checkerboard and stripe quantum solid phases at half filling factor. Many of these solid phases along with multiple supersolid regimes were numerically observed by means of quantum Monte Carlo (QMC) in our group~\cite{PhysRevA.103.043333,Zhang2022}.

\begin{figure}[!h]
\centering
\includegraphics[width=0.9\linewidth]{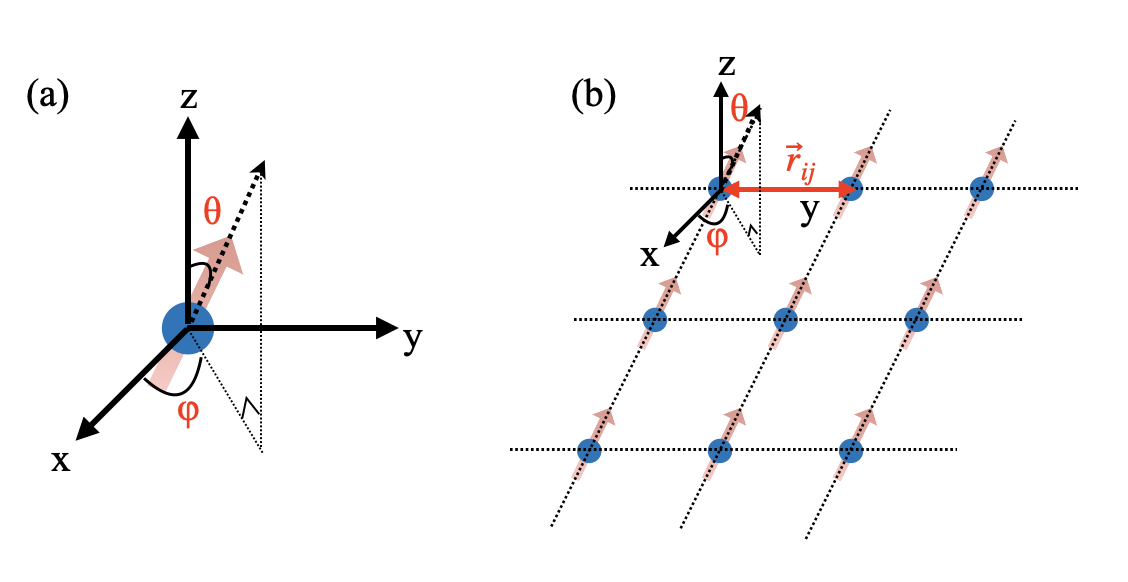}
\caption{Schematic representation of the system. Dipoles are trapped in a two-dimensional optical lattice and are aligned parallel to each other along the direction of polarization. $\theta$ is the polar angle between
polarization axis and z direction; $\varphi$ is the azimuthal angle, in this work we fix $\varphi=45^\circ$. $\vec {r}_{ij}= (x_{ij},y_{ij})$ is the relative position between site $i$ and $j$. }
\label{setup}
\end{figure}

The work presented here extends our previous studies and is motivated by the findings of Ref.~\cite{Su2023}. In this manuscript, we consider a fixed azimuthal angle $\varphi = 45^\circ$ and employ path-integral quantum Monte Carlo simulations using the worm algorithm~\cite{PROKOFEV1998253} to investigate density instabilities. Such instabilities arise due to first order phase transitions appearing for polar angle $\theta\gtrsim \theta_i $ with $\theta_i(V/J)\in [62^\circ,68^\circ]$ at zero temperature. Finite temperature shifts the value of $\theta_i$ towards larger values, and stabilizes density values otherwise unstable at lower temperatures into a phase-separated state.  In this work, we treat the polar angle $\theta$ and the dipolar interaction strength as independent control parameters (see figure~\ref{setup}). We first focus on half-filling factor and investigate which phases are stabilized at zero-temperature. We find that for weak interactions, the system remains in the superfluid (SF) phase for all polar angles. As the interaction strength increases, the SF phase becomes unstable and gives way to either a checkerboard solid (CB) at small polar angles or a double diagonal stripe solid (double-DSS) at intermediate polar angles. For 
lower interaction strengths, the CB and double-DSS phases are separated by an intervening SF region.
The double-DSS phase is characterized by alternating occupied and empty diagonal stripes, with the stripe width of two lattice sites. Upon further increasing the polar angle, the system remains in an incompressible phase (IP) where particles organize into diagonal stripes of varying thickness, separated by empty stripes whose widths also vary. The specifics of the stripe arrangement depends on system size and initial conditions. 

For sufficiently large polar angles, half-filling becomes {\textit {unstable}}. Instead, the system undergoes a first-order phase transition between  a completely empty lattice (zero filling) and a fully occupied lattice (unit filling). This behavior differs from the findings reported in~\cite{Su2023}, where a self-bound insulating phase, i.e. a phase with lattice separated into
unity-filled and empty regions each occupying only half of the sites,  was observed at half-filling for polar angle $\theta\gtrsim 80^\circ $. Our simulations show that at these polar angles, the ground state of the homogeneous system does not support a self-bound insulator. Notably, though, phase separation into unit-filled and empty regions at large polar angles becomes stable at {\it{finite}} temperature.  Our finite-temperature simulations indicate that fillings unstable at zero temperature can be stabilized as the temperature increases.  In this regime, the particles spontaneously phase-separate into a region of unit filling coexisting with an empty region. A separation between fully filled and empty regions, can also be observed when the system is constrained to an otherwise unstable number of particles. As expected, this results in coexistence of the two stable phases -- fully filled and empty lattice regions -- otherwise separated by a first order phase transition.  Finally, we considered the case of an harmonically trapped system and found that density instabilities still exists in this non-homogeneous setup, and that constraining the number of particles to an otherwise thermodynamically unstable value results in a particle configuration which resembles a self-bound insulator localized at the trap center.

This paper is organized as follows: In Section 2, we discuss the Hamiltonian describing the system. In Section 3, we present the phases stabilized at half-filling and zero temperature. In Section 4, we analyze the effect of finite temperature on density instabilities and explore the effects of the harmonic confinement. In Section 5, we conclude the paper.

\begin{figure}[!h]
    \centering
    \includegraphics[width=0.9\linewidth]{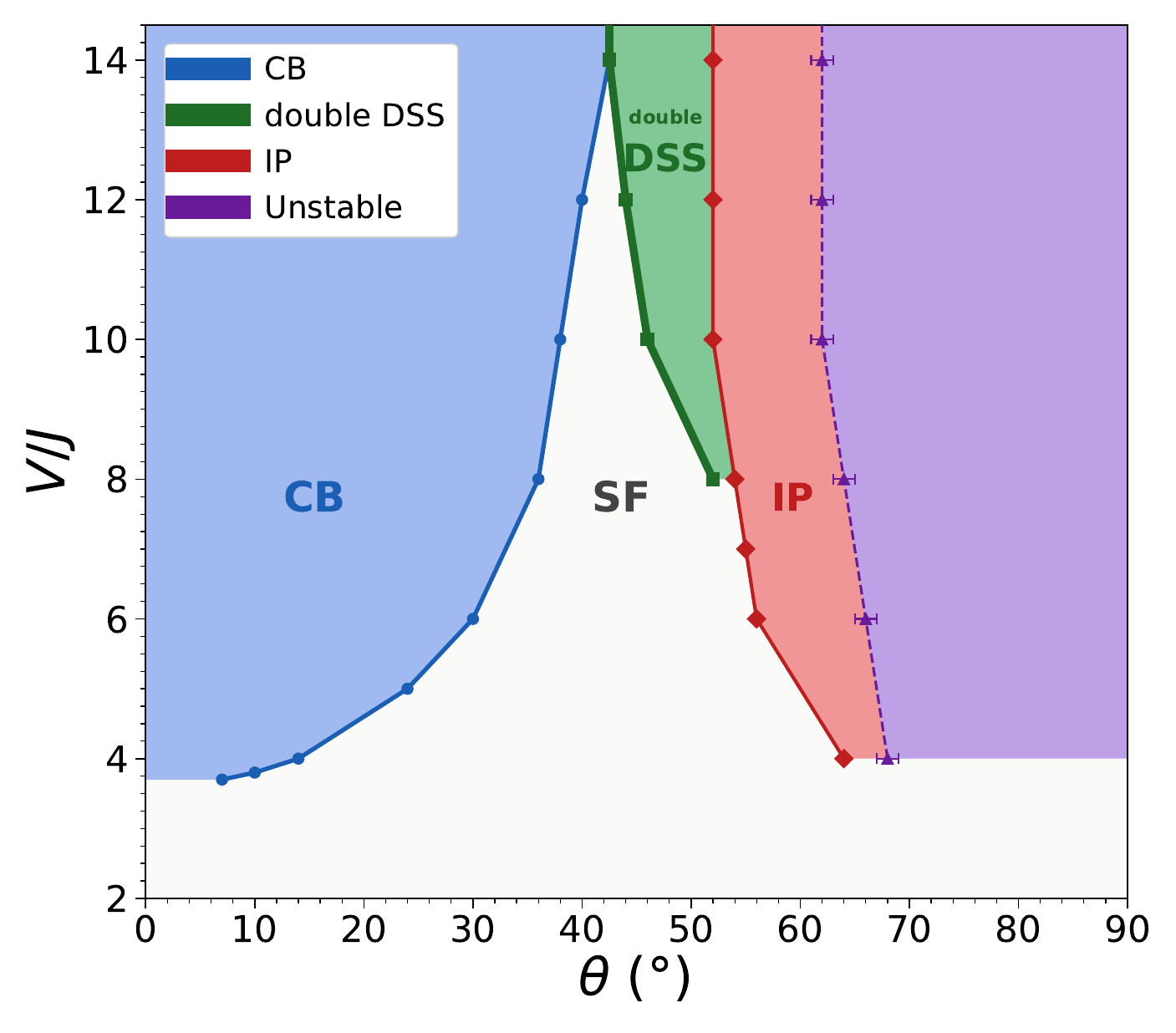}
    \caption{Zero-temperature phase diagram of the homogeneous system described by Eq.~(\ref{eqn1}) as a function of polar angle \( \theta \) and \( V/J \), at half-filling and fixed azimuthal angle \( \varphi = 45^\circ \). The system features a superfluid phase (SF), shown in white; a checkerboard solid (CB), blue region; a double diagonal stripe solid (double-DSS), green region; incompressible region (IP), red region where particles tend to organize into stripe of increasing thickness as we increase $\theta$. Thick solid lines indicate first-order phase transitions; the thickness of the blue line separating the CB phase from the SF phase is $\delta \theta=0.5 ^{\circ}$, while the thickness of the green line separating the SF phase from the double-DSS is $\delta \theta=1 ^{\circ}$, and the the thickness of the red lines separating the SF and double-DSS phases from the IP region is $\delta \theta=0.3 ^{\circ}$ The purple area corresponds to a region of instability for any density \( n \neq 0, 1 \). In this parameter region, the system is either empty (\( n = 0 \)) or fully occupied (\( n = 1 \)), undergoing a first-order phase transition between empty and full states as a function of chemical potential \( \mu \) at fixed \( V/J \) and \( \theta \).}
    \label{phase}
    \end{figure}

\begin{figure*}[ht]
 \centering
    \includegraphics[width=1\linewidth]{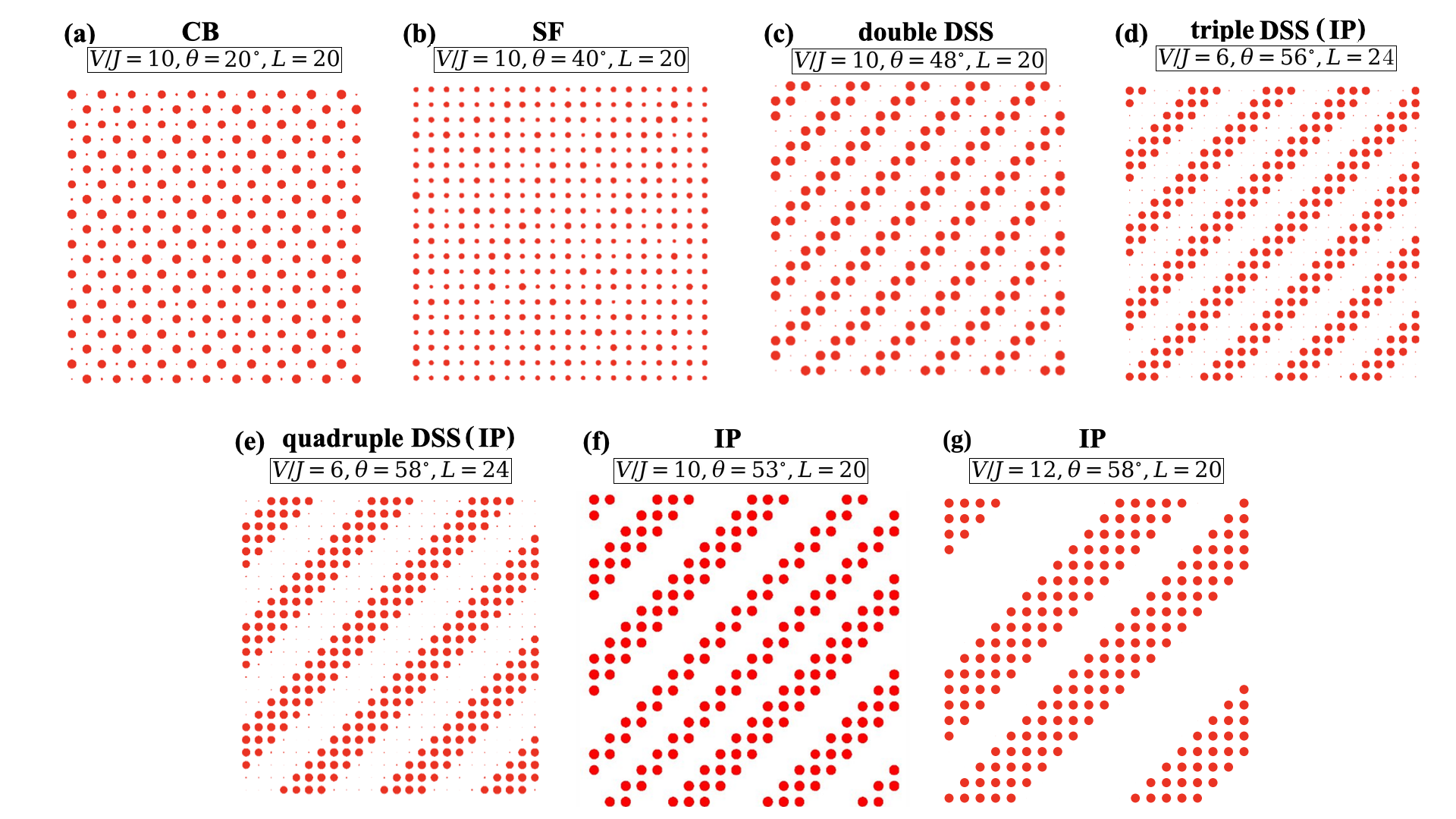}
\caption{Density maps representative of the distinct phases identified 
in the phase diagram (Fig.~\ref{phase}). Each red circle corresponds to a 
particle and its radius is proportional to the local occupation 
density $\langle n_i \rangle$, where the average is taken over a single Monte Carlo configuration. For all panels $\beta = L$; the remaining 
parameters are indicated in the box above each plot. 
(a)~Checkerboard solid (CB): particles 
arrange in a staggered pattern with alternating occupied and empty sites, 
breaking translational symmetry with a periodicity of two lattice constants along both directions. 
(b)~Superfluid (SF): the density is 
uniform across all sites. 
(c)~Double diagonal stripe solid (double DSS): 
the occupied sites organise into pairs of parallel diagonal stripes 
separated by empty ones.
(d)~Triple diagonal stripe solid (triple DSS): 
similar to the double DSS but with stripes of thickness of three sites.
(e)~Quadruple diagonal stripe solid: similar to the double DSS but with stripes of thickness of four sites. 
Importantly, the triple and quadruple DSS phases exist only over a 
very narrow range of tilt angles and are difficult to stabilise: they 
do not appear consistently independently on the initial conditions, indicating metastability due to the presence of competing local minima of energy. 
(f)-(g)~Incompressible phase (IP): with diagonal stripe of different widths. }
    \label{fig2}
\end{figure*}

\section{Hamiltonian}

We consider a system of hardcore dipolar bosons confined to a square optical lattice with lattice constant $a$, which we take as the unit of length. The hardcore constraint is motivated by experimental realizations~\cite{Su2023}, where the on-site interaction energy is much larger than all other relevant energy scales, effectively prohibiting double occupancy. The dipoles are aligned parallel to one another along the direction of polarization, each carrying an induced dipole moment $d$. This polarization direction can be tuned experimentally using an external magnetic field. Here, we consider this direction to be laying in a plane perpendicular to the lattice plane at an angle of $ 45^\circ$ with the $x$-axis. Accordingly, throughout this work we fix the azimuthal angle to $ \varphi=45^\circ$ and vary the polar angle $\theta$, see figure~\ref{setup}.

In the framework of second quantization and using the Wannier basis \cite{Kohn1959}, the Hamiltonian describing two-dimensional dipolar bosons trapped in an optical lattice can be expressed, for the lowest Bloch band, as follows:

\begin{multline}
H = -J \sum_{\langle i,j \rangle} {a}_i^\dagger {a}_j - \sum_i \mu_i n_i
\\
+V \sum_{i<j} \frac{n_i n_j [r_{ij}^2 - 3 \sin^2\theta(x_{ij}\cos\varphi+y_{ij}\sin\varphi)^2]}{r_{ij}^5} .
\label{eqn1}
\end{multline}
Here, $a_i^\dagger$ ($a_i$) are the bosonic creation (annihilation) operators with the usual commutation relations and satisfying the hard-core condition $a_i^{\dagger 2}=0$, and $n_i=a_i^\dagger a_i$.  We use $\langle \dots \rangle$ to denote nearest neighboring sites. In Eq.~\ref{eqn1}, the first term describes the kinetic energy of the system with hopping energy $J$; the second term 
$\mu_i=\mu-W_i$ is the sum of the chemical potential $\mu$ which sets the total number of particles and the confining potential $W_i$; the last term is the dipole-dipole interaction with  $V\propto d^2/a^3$ the strength of the nearest-neighbor repulsive interaction energy when the dipoles are oriented perpendicular to the square lattice, $r_{ij}=\vert \vec {r}_i-\vec {r}_j\vert=\vert (x_{ij},y_{ij})\vert$ is the relative distance between site $i$ and site $j$, $\theta$ and $\varphi$ are the polar and azimuthal angle, respectively. In this work, we fix $\varphi=45^\circ$, and vary $\theta$ and $V$. Unless otherwise specified, we set $W_i=0$, set periodic boundary conditions in space, and tune the chemical potential to achieve half-filling. We set the interaction cutoff distance to L, the system size.

\section{Quantum Phases and density instabilities at T=0}
\label{sec:sec3}

\begin{figure} [h]
    \centering
       \includegraphics[width=0.9\linewidth]{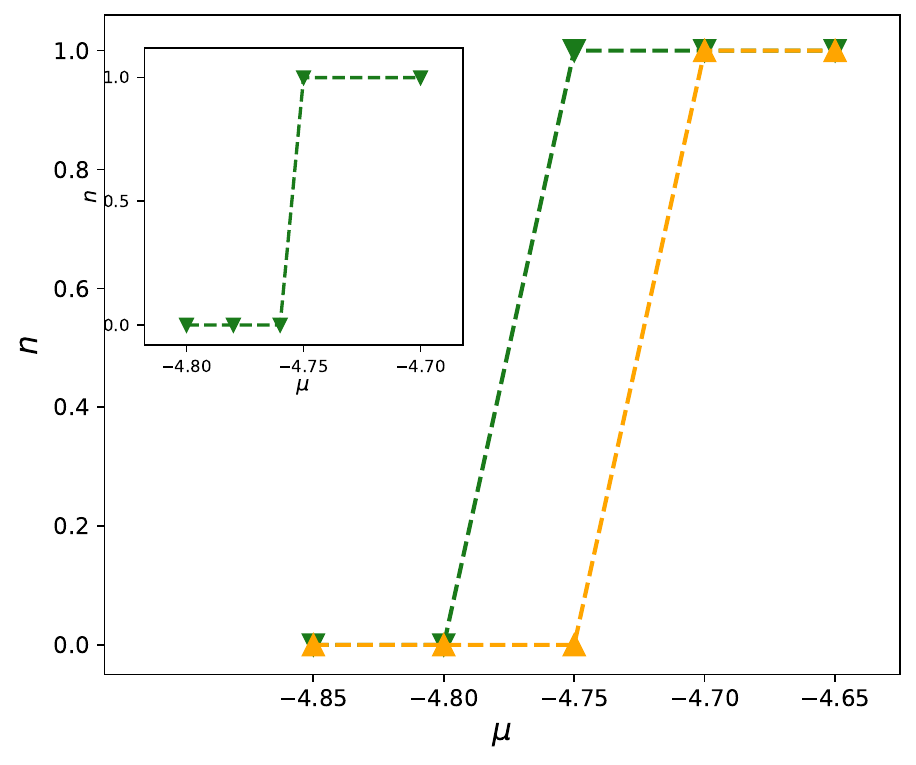}
       \caption{ Main plot: Representative hysteretic behavior of $n$ obtained by sweeping $\mu$ in forward (increasing $\mu$, orange triangles $\triangle$) and backward (decreasing $\mu$, green triangles $\triangledown$) directions, where the system is initialized at each value of $\mu$ with the equilibrium configuration from the previous step. 
       Inset: filling factor $n$ as a function of chemical potential $\mu$, showing an abrupt jump from $n = 0$ to $n = 1$, with no stable intermediate filling accessible in this regime.}

    \label{hysteresis}
\end{figure}
\label{sec:sec5}

\label{sec:sec4}
\begin{figure*} [!ht]
    \centering
       \includegraphics[width=1.0\linewidth]{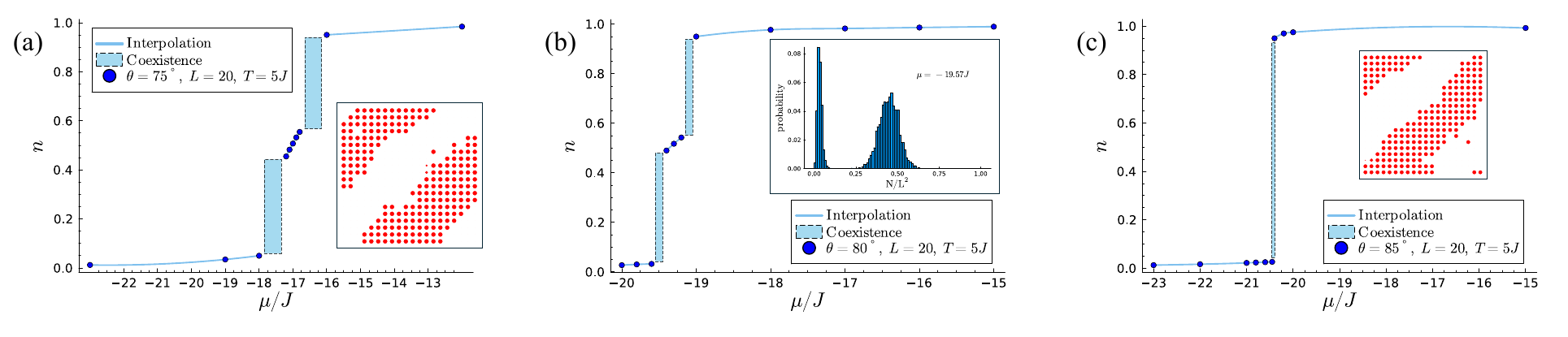}
       \caption{Main plots: $n$ as function of  $\mu$ for $\theta=75^{\circ}$ (a), $\theta=80^{\circ}$ (b) and $\theta=85^{\circ}$ (c) at $V/J=10$ and temperature $k_B T=5J$. Solid dots denote thermodynamically stable values of $n$ for a given $\mu$, while the light blue shaded regions indicate intervals of $\mu$ where phase coexistence occurs.
    Insets: (a) representative density map of the stable phase-separated state obtained for $\mu=-17J$. (b) Bimodal histogram of the filling factor observed for $\mu=-19.57J$ which signals coexistence for this $\mu$.
    (c) Representative density map corresponding to phase coexistence between empty and fully-filled regions when the simulation is constrained to filling different than the stable values. }
    \label{fig:T3_nprofiles}
\end{figure*}
In this Section, we present the ground state phase diagram of Hamiltonian~\ref{eqn1} with $W_i=0$ in the $V/J$-$\theta$ plane at fixed filling factor $n=0.5$ (figure~\ref{phase}). We considered system sizes in the range $12 \leq L \leq 48$ and an inverse temperature $\beta=L/J$ to ensure the system is effectively at zero temperature.

Four distinct phases are observed in the system: the superfluid, checkerboard solid, double diagonal stripe solid, and incompressible phases. The superfluid phase is characterized by off-diagonal long-range order and a finite superfluid stiffness $\rho_s$. In this phase, the compressibility $\kappa$ is also finite. The superfluid stiffness can be computed from the statistics of the winding number in space as $\rho_{s}=\frac{\langle \mathbf{W}^2\rangle}{DL^{D-2}\beta}$, where $\langle\mathbf{W}^2\rangle=\sum_{i=1}^D \langle W_i^2\rangle$ is the expectation value of the squared winding number, $D$ is the spatial dimension of the system (here $D=2$), $L$ is the linear system size, and $\beta$ is the inverse temperature~\cite{Winding}. The compressibility can be evaluated from the fluctuations of the particle number in imaginary time as $\kappa=\frac{\beta \Delta N^2}{L^2}$, where $\Delta N ^2=\langle(N-\langle N\rangle)^2\rangle$ with $N$ the total number of particles. In the solid phases, long-range diagonal order is signaled by a finite structure factor $S(\mathbf{k})=\sum_{\mathbf{r,r'}} \exp[i \mathbf{k (\mathbf{r-r'})}]\langle n_{\mathbf{r}} n_{\mathbf{r'}} \rangle /N^2$, where $\mathbf{k}$ is the corresponding reciprocal lattice vector. For instance, in the CB solid, $\mathbf{k}=(\pi, \pi)$ while in the double-DSS, $\mathbf{k}=(\pi/2, \pi/2)$. Solid phases are incompressible, with $\kappa=0$

Fig.~\ref{phase} shows that at
lower interaction strengths ($V/J \le 3.7$), the system remains in a SF phase regardless of the tilt angle $\theta$. For $\theta \lesssim 42^\circ$, there exists a $\theta$-dependent value of $V/J$ at which the SF phase gives way to a CB solid (blue region in Fig.~\ref{phase}) via a first-order phase transition as evidenced by abrupt jumps in the $\rho_s$ and $S(\pi,\pi)$. The first order transition is indicated by the thick solid blue line. The position of the first-order transition is within a range $\delta\theta \sim 0.5^\circ$ which corresponds to the width of the hysteresis observed in $\rho_s$ and $S(\mathbf{k})$ as functions of $V/J$ or $\theta$. 
As expected, the SF-CB first-order boundary shifts toward larger values of $V/J$ with increasing polar angle, reflecting the fact that increasing $\theta$ at fixed $V/J$ reduces the nearest-neighbor repulsive interaction in favor of an increasing attractive interaction along the diagonal.

For polar angles in the range $42^\circ \lesssim \theta \lesssim 52^\circ$, the SF phase becomes unstable toward a double-DSS (green region in Fig.~\ref{phase}). The double-DSS is a diagonal solid characterized by stripes two sites thick. The transition from SF to double-DSS is also a first-order transition, represented by the green line in Fig.~\ref{phase}. The thickness $\delta\theta \sim 1^\circ$ of this line similarly reflects the hysteresis width of $\rho_s$ and $S(\mathbf{k})$ as a function of $\theta$. Figure~\ref{fig2} shows representative  density maps of CB (a), SF (b), double-DSS (c) at $V/J=10$ for a system of size $L=20$. Here, each circle corresponds to particle, and its radius is proportional to the local density averaged over a single QMC configuration. 

We notice that, up to $V/J\sim 14$, the SF phase remains stable over a finite range of $\theta$, intervening between the CB and double-DSS phases. This behavior arises from the competition between the repulsive nearest-neighbor interaction and the attractive interaction along the lattice diagonals at intermediate angles.  For sufficiently strong dipolar interactions ($V/J\gtrsim 14$), the SF phase disappears, and a first-order phase transition between the CB and double-DSS phases occurs at $\theta\sim42^{\circ}$.

As the polar angle increases, the system remains in an incompressible phase; however, the double-DSS phase is no longer energetically favored. Instead, particles tend to organize into stripe of increasing thickness (IP phase). Owing to the finite system sizes in our simulations, we do not observe a unique solid ordering for fixed values of $V/J$, $\theta$. Rather, the resulting particle arrangement depends on the system size $L$. In this region of parameter space, we also observed that different initial conditions may result in slightly different particle arrangement indicating the presence of metastability. For example, in the range $6 \lesssim V/J \lesssim 8$, a triple-DSS phase is consistently observed for system sizes with $L$ being a multiple of 6, Fig.~\ref{fig2} (d). For stronger dipolar interactions, however, identifying a well-defined triple-DSS configuration becomes increasingly difficult. Similar considerations apply to the quadruple-DSS, Fig.~\ref{fig2} (e), which is challenging to observe even in the range $6 \lesssim V/J \lesssim 8$. 
More generally, throughout the parameter regime labeled IP in Fig.~\ref{fig2}, no unique stripe ordering emerges for fixed $V/J$ and $\theta$. Instead, particles form diagonal stripes of varying thickness, separated by empty stripes whose widths also vary, depending on system size and initial conditions. Increasing the polar angle favors the formation of thicker stripes, including the (finite-size) system being separated in two regions only -- a unit filled stripe and an empty stripe each of thickness $L/2$. These observations suggest the existence of many nearly degenerate local minima, separated by substantial energy barriers, within this region of the phase diagram. As a consequence, reaching the true ground state at effectively zero temperature is expected to be experimentally challenging. Furthermore, since experiments are necessarily performed on finite systems —comparable in size to those considered in our simulations— finite-size effects are also expected to play an important role in determining the observed particle arrangement.


 
For $\theta\gtrsim \theta_i $ with $\theta_i\in [62^\circ,68^\circ]$, half-filling becomes {\it{unstable}} (purple region in figure~\ref{phase}). Within this region of the parameter space, the only stable filling factors are either zero or unity, depending on the chemical potential, with a first order phase transition separating these two stable values of $n$. Figure~\ref{hysteresis} shows the abrupt jump of $n$ from zero to one upon increasing $\mu$ (inset), as well as the hysteretic behavior of the density obtained by varying $\mu$ in forward and backward sweeps. The latter is performed by initializing the system at each value of $\mu$ with the equilibrium configuration corresponding to the previous value in the sweep. This behavior is consistently observed—independent of initial conditions—for all system sizes explored, and for all polar angles within this region.

\section{INSTABILITY REGION}

In Reference~\cite{Su2023}, 
a self-bound insulator—i.e., a globally phase-separated state in which the lattice splits into unity-filled and empty regions—is experimentally observed for polar angles $\theta \gtrsim80^\circ$. These observations are made at $V/J\simeq 10$ and an estimated experimental temperature $k_BT\simeq0.5V=5J$~\cite{Su2023}. While, in absolute terms, for the experimental setup in~\cite{Su2023}, this temperature corresponds to only several hundreds picokelvin, our results reveal that for polar angles $\theta\gtrsim \theta_i $ with $\theta_i\in [62^\circ,68^\circ]$, half-filling is not stable in the ground state of a homogeneous system for all system sizes considered  ($10\leq L\leq 42 $).  As a consequence, it appears that a self-bound insulator is not spontaneously stabilized by dipolar interaction in the bulk ground state and that, certainly in this parameter range, a temperature $k_BT\simeq5J$ cannot be considered sufficiently low to probe ground-state behavior. Therefore, we study the role of temperature and the fate of density instabilities at finite temperature in order to understand under which conditions a self-bound insulator can be stabilized.

In the following, we present our simulation results at finite T, focusing on $k_BT=5J$. We also performed simulations at $k_BT=J, 2J, 3J$, and obtained qualitatively similar results.  Overall, we find that the onset of density instabilities shifts to larger polar angles as temperature increases. In addition, thermal fluctuations stabilize intermediate filling factors around half-filling. The range of these temperature-stabilized  intermediate filling factors broadens with increasing T. For all system sizes studied ($12\leq L\leq 48$), the equilibrium state at these intermediate fillings corresponds to phase separation between empty and fully filled regions, separated diagonally, with the fraction of occupied  sites determined by the overall filling factor. This is consistent with the appearance of increasingly wider stripe-like structures upon increasing polar angle in the metastable region of the ground-state phase diagram, including a phase-separated state in the proximity of the onset of the first order phase transition.

In Fig.~\ref{fig:T3_nprofiles}, we show the density $n$ as a function of the chemical potential $\mu$ for $\theta = 75^\circ $ (a), $80^\circ$ (b), and $85^\circ$ (c) at $V/J=10$ and temperature $k_BT = 5J$. Solid dots denote thermodynamically stable values of $n$ for a given $\mu$, while the light-blue shaded regions indicate intervals of $\mu$ where phase coexistence occurs.
We note that for these same  polar angles, the ground state only supports the fillings $n = 0$ and $n = 1$. At finite temperature, however, a finite interval of densities around $n = 0.5$ becomes stable and correspond to a thermodynamically stable phase-separated state (see inset of Fig.~\ref{fig:T3_nprofiles} (a) where we show a representative density map of the phase-separated state). First-order phase transitions, marked by the shaded regions, are still present. Within these regions, the simulation samples both competing equilibrium states, corresponding to the two densities that bound the coexistence interval, indicating the presence of competing  free-energy minima. This is illustrated in the inset of Fig.~\ref{fig:T3_nprofiles}(b), where an example of a bimodal histogram of the filling factor is shown. 
At $\theta = 85^\circ$, we recover the ground-state behavior, with stability restricted to $n = 0$ and $n = 1$, Fig.~\ref{fig:T3_nprofiles}(c). If the simulation is constrained to fillings different from these stable values, phase coexistence between the empty and fully filled phases is observed, as shown in the inset of Fig.~\ref{fig:T3_nprofiles}(c), where a representative density map is displayed. 

\begin{figure} [h]
    \centering
       \includegraphics[width=1\linewidth]{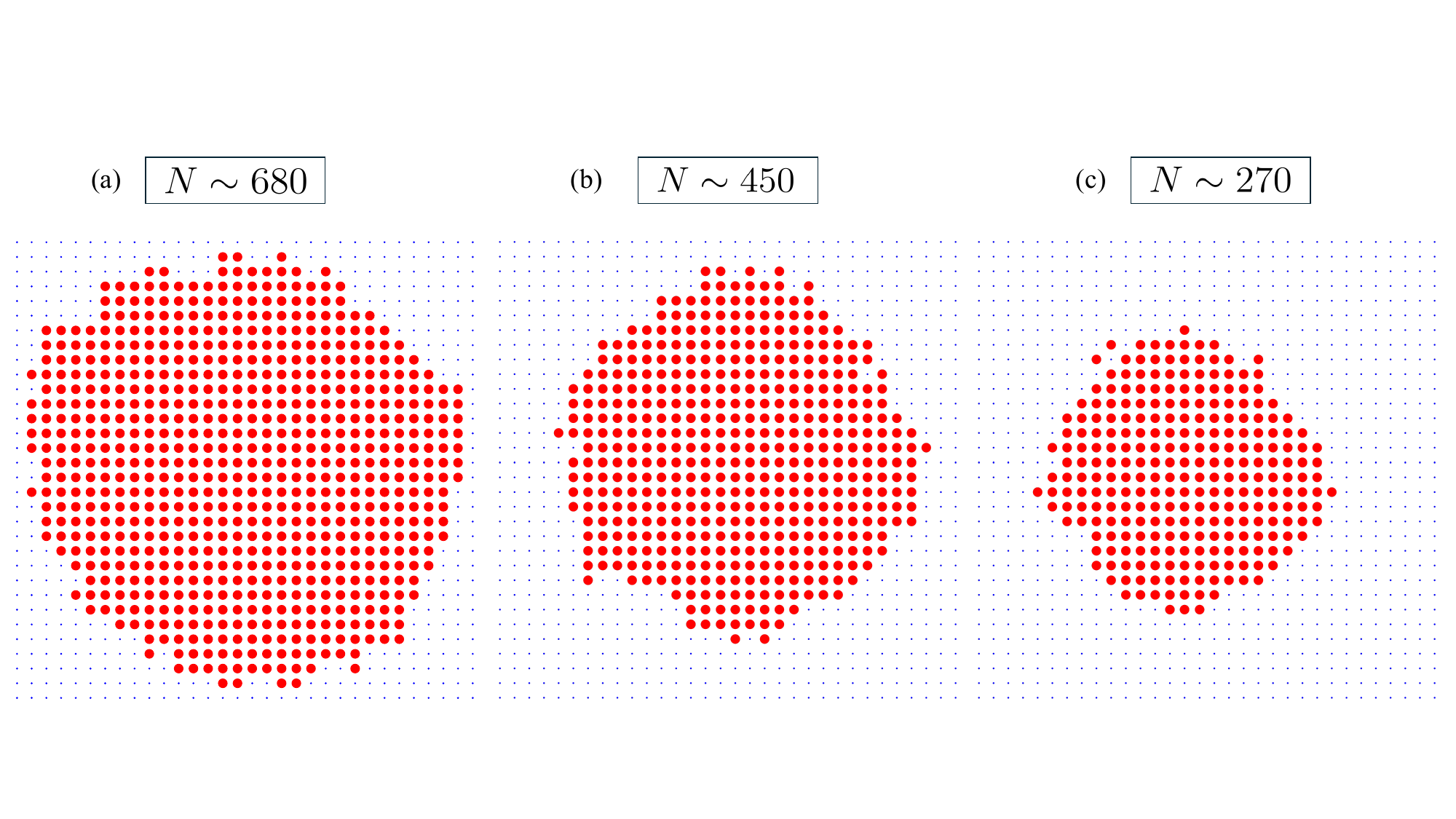}
       \caption{Density maps in the presence of an external  harmonic trap: $L=30$, $k_BT=5J$, $V=10J$, $\theta=80^\circ$, $W=0.003J$. The radius of the red circle is proportional to the local occupation density $\langle n_i \rangle$, where the average is taken over a single Monte Carlo simulation. The blue dots depict empty lattice sites. (a) Stable phase with $N\sim 680$. Particles occupy the central region of the trap, while the outer region remains essentially empty. This particle configuration is the counterpart, in the trap, of the phase separation state in the homogeneous system. (b)-(c) Density map obtained when the particle number is constrained to an otherwise thermodynamically unstable value, $N \sim 450$ (b), $N \sim270$ (c). These particle configurations resemble a self-bound insulator localized at the center of the trap.}

    \label{fig6: parabolic trap}
\end{figure}

Finally, we performed simulations in the presence of a harmonic trapping potential. For these simulations, we used hard-wall boundary conditions. We considered system parameters $L=30$, $k_BT=5J$, $V=10J$, $\theta=80^\circ$, and trap strength $W=0.003J$. As the trap breaks translational invariance, one might expect signatures of first order phase transition to appear as abrupt changes in the density profile as a function of the distance from the trap center. Instead, for the parameters considered here, we still observe a discontinuous change in the total particle number as the chemical potential is varied.
As $\mu$ is increased, the trapped system undergoes a transition from an empty state to a state containing approximately $N\sim680$ particles. In the latter state, particles occupy the central region of the trap, while the outer region remains essentially empty (Fig. \ref{fig6: parabolic trap}(a)). This configuration is the counterpart, in the trap, of the phase separation state in the homogeneous system. Due to the presence of the external potential, the local chemical potential is maximum at the center of the trap and correspondingly, the unit-filled region is the one at the center of the trap.  If the particle number is constrained to lower (unstable) values, the resulting particle configuration also resembles a self-bound insulator localized at the trap center (Fig. \ref{fig6: parabolic trap}(b)-(c)). However, in this case, this configuration does not correspond to a thermodynamically stable self-bound insulator. Rather, it originates from phase coexistence associated with the underlying first-order transition discussed above.

Overall, our finite-temperature simulations show that thermal fluctuations stabilize intermediate filling factors at polar angles $\theta$ for which the ground state is unstable for all $n\neq 0,1$. While the first-order nature of the transition is preserved, thermal fluctuations give rise to a thermodynamically stable phase-separated state at intermediate densities around half-filling. As $\theta$ increases, the interval of stable filling factors between $n\sim0$ and $n\sim1$ progressively narrows, eventually collapsing to a small region centered around $n \approx 0.5$ (in addition to the stable states at $n\sim0$ and $n\sim1$). Upon further increasing $\theta$, the equilibrium states at intermediate density disappear altogether, and the system undergoes a direct first-order transition between the $n\sim0$ and $n\sim1$, analogous to the first order phase transition observed in the  ground state. 
Taken together with our ground-state results, these findings indicate that the self-bound insulator observed experimentally  relies on finite-temperature effects that stabilize half-filling as a phase-separated equilibrium state in the homogeneous system.  Realizing a self-bound insulator at half-filling in the experimental 'region of interest' at the center of the trap therefore requires a very careful choice of Hamiltonian parameters.
We further showed that density configurations resembling a self-bound insulator may also arise at the center of a harmonic trap as a consequence of phase coexistence associated with a first-order transition when the particle number is constrained to values that are thermodynamically unstable in equilibrium. Such configurations should therefore be distinguished from a thermodynamically stable self-bound insulating phase. In this respect, QMC benchmarking is essential for interpreting experimental observations and for establishing whether an observed self-bound structure corresponds to an equilibrium phase or to phase coexistence.

\section{CONCLUSIONS}

We have investigated the zero-temperature properties and density instabilities of hard-core dipolar bosons on a two-dimensional square lattice at fixed azimuthal angle $\varphi=45^\circ$ using path-integral quantum Monte Carlo simulations with the worm algorithm. At half filling, the ground-state phase diagram exhibits superfluid, checkerboard solid, double-diagonal stripe solid, and incompressible phases. For polar angles $\theta\gtrsim \theta_i$, there exists a density instability in which only the empty ($n=0$) and fully filled ($n=1$) states are thermodynamically stable. In this regime, intermediate fillings are excluded by a first-order transition, and a homogeneous self-bound insulating phase at half filling is absent.
At finite temperature, thermal fluctuations shift the onset of density instabilities to larger polar angles and stabilize a finite range of intermediate fillings around $n\approx0.5$. The corresponding equilibrium states are phase-separated configurations composed of one empty and one fully filled diagonal domain. These configurations closely resemble the self-bound insulator observed in recent experiments~\cite{Su2023}. In the presence of a harmonic trap, phase-separated particle configurations similar to the self-bound structure also appear. They can also emerge at the trap center when the particle number lies within the coexistence region of the underlying first-order transition.
Overall, our results indicate that the experimentally observed self-bound insulator is not a zero-temperature phase of the homogeneous system, but rather a finite-temperature phase-separated state stabilized by thermal fluctuations. Moreover, our work shows that, in certain parameter regimes, density configurations alone may not be sufficient to distinguish genuine equilibrium phases from phase coexistence. In this respect, unbiased quantum Monte Carlo benchmarks are essential for interpreting experimental observations. Finally, the strong sensitivity of density instabilities and phase-separated states to temperature~\cite{Lingua:2017aa} and dipole orientation suggests new opportunities for thermometry in dipolar quantum simulators.

\label{sec:sec5}

\begin{acknowledgments}
C. Zhang acknowledges support from the National Natural Science Foundation of China (NSFC) under Grants No. 12204173 and 12275002, the University Annual Scientific Research Plan of Anhui Province under Grant No. 2022AH010013, and the Education Department of Anhui Province. The computing for this project was performed at the cluster at Clark University and Politecnico di Torino.
\end{acknowledgments}

\bibliography{references} 

@article{Lingua:2017aa,
	author = {Lingua, F. and Capogrosso-Sansone, B. and Minardi, F. and Penna, V.},
	date = {2017/07/11},
	doi = {10.1038/s41598-017-05353-6},
	id = {Lingua2017},
	isbn = {2045-2322},
	journal = {Scientific Reports},
	number = {1},
	pages = {5105},
	title = {Thermometry of bosonic mixtures in Optical Lattices via Demixing},
	url = {https://doi.org/10.1038/s41598-017-05353-6},
	volume = {7},
	year = {2017}}

@article{RecatiStringari2023,
  author  = {Recati, Alessio and Stringari, Sandro},
  title   = {Supersolid behavior of dipolar Bose gases},
  journal = {Nature Reviews Physics},
  year    = {2023},
  volume  = {5},
  pages   = {735--747},
  doi     = {10.1038/s42254-023-00648-2}
}

@article{Mukherjee2023,
  author  = {Mukherjee, B. and others},
  title   = {Quantum droplets and supersolidity in dipolar Bose gases},
  journal = {Comptes Rendus Physique},
  year    = {2023},
  volume  = {24},
  pages   = {1--28},
  doi     = {10.5802/crphys.154}
}

@article{Zampronio2024,
  author  = {Zampronio, S. and Macr{\`i}, T. and Pohl, T.},
  title   = {Quasicrystalline supersolids in dipolar quantum gases},
  journal = {Physical Review Letters},
  year    = {2024},
  volume  = {132},
  pages   = {123401},
  doi     = {10.1103/PhysRevLett.132.123401}
}

@article{SOCdipolar2024,
  author  = {Zhang, Y. and others},
  title   = {Topological phases in spin--orbit-coupled dipolar Bose gases},
  journal = {Physical Review A},
  year    = {2024},
  volume  = {109},
  pages   = {013303},
  doi     = {10.1103/PhysRevA.109.013303}
}

@article{He2025,
  author  = {He, L. and others},
  title   = {Low-dimensional dipolar quantum gases: recent advances},
  journal = {Reports on Progress in Physics},
  year    = {2025},
  volume  = {88},
  pages   = {012401},
  doi     = {10.1088/1361-6633/adxxxx}
}

@article{PhysRevLett.105.135301,
  title = {Supersolid Droplet Crystal in a Dipole-Blockaded Gas},
  author = {Cinti, F. and Jain, P. and Boninsegni, M. and Micheli, A. and Zoller, P. and Pupillo, G.},
  journal = {Phys. Rev. Lett.},
  volume = {105},
  issue = {13},
  pages = {135301},
  numpages = {4},
  year = {2010},
  month = {Sep},
  publisher = {American Physical Society},
  doi = {10.1103/PhysRevLett.105.135301},
  url = {https://link.aps.org/doi/10.1103/PhysRevLett.105.135301}
}

@article{PhysRevB.87.081106,
  title = {Topological phases in ultracold polar-molecule quantum magnets},
  author = {Manmana, Salvatore R. and Stoudenmire, E. M. and Hazzard, Kaden R. A. and Rey, Ana Maria and Gorshkov, Alexey V.},
  journal = {Phys. Rev. B},
  volume = {87},
  issue = {8},
  pages = {081106},
  numpages = {6},
  year = {2013},
  month = {Feb},
  publisher = {American Physical Society},
  doi = {10.1103/PhysRevB.87.081106},
  url = {https://link.aps.org/doi/10.1103/PhysRevB.87.081106}
}

@article{lhfx-c4xr,
  title = {Topological quantum floating phase of dipolar bosons in an optical ladder},
  author = {Korbmacher, Henning and Dom\'{\i}nguez-Castro, Gustavo A. and \L{}\k{a}cki, Mateusz and Zakrzewski, Jakub and Santos, Luis},
  journal = {Phys. Rev. A},
  volume = {112},
  issue = {1},
  pages = {L011301},
  numpages = {6},
  year = {2025},
  month = {Jul},
  publisher = {American Physical Society},
  doi = {10.1103/lhfx-c4xr},
  url = {https://link.aps.org/doi/10.1103/lhfx-c4xr}
}

@article{Chomaz_2023,
doi = {10.1088/1361-6633/aca814},
url = {https://doi.org/10.1088/1361-6633/aca814},
year = {2022},
month = {dec},
publisher = {IOP Publishing},
volume = {86},
number = {2},
pages = {026401},
author = {Chomaz, Lauriane and Ferrier-Barbut, Igor and Ferlaino, Francesca and Laburthe-Tolra, Bruno and Lev, Benjamin L and Pfau, Tilman},
title = {Dipolar physics: a review of experiments with magnetic quantum gases},
journal = {Reports on Progress in Physics}
}

@article{Langen:2024aa,
	author = {Langen, Tim and Valtolina, Giacomo and Wang, Dajun and Ye, Jun},
	date = {2024/05/01},
	doi = {10.1038/s41567-024-02423-1},
	id = {Langen2024},
	isbn = {1745-2481},
	journal = {Nature Physics},
	number = {5},
	pages = {702--712},
	title = {Quantum state manipulation and cooling of ultracold molecules},
	url = {https://doi.org/10.1038/s41567-024-02423-1},
	volume = {20},
	year = {2024}}

@ARTICLE{Sinha_supersolids,
       author = {{Sinha}, Sudip and {Sinha}, Subhasis},
        title = "{Supersolid phases of bosons}",
      journal = {Journal of Physics Condensed Matter},
         year = 2025,
        month = aug,
       volume = {37},
       number = {33},
          eid = {333001},
        pages = {333001},
          doi = {10.1088/1361-648X/adf6fb},
archivePrefix = {arXiv},
       eprint = {2502.06660},
 primaryClass = {cond-mat.quant-gas},
       adsurl = {https://ui.adsabs.harvard.edu/abs/2025JPCM...37G3001S}
}

@article{PhysRevB.111.024511,
  title = {Dipolar bosons in a twisted bilayer geometry},
  author = {Zhang, Chao and Fan, Zhijie and Capogrosso-Sansone, Barbara and Deng, Youjin},
  journal = {Phys. Rev. B},
  volume = {111},
  issue = {2},
  pages = {024511},
  numpages = {7},
  year = {2025},
  month = {Jan},
  publisher = {American Physical Society},
  doi = {10.1103/PhysRevB.111.024511},
  url = {https://link.aps.org/doi/10.1103/PhysRevB.111.024511}
}

@article{PhysRevA.97.013615,
  title = {Equilibrium phases of dipolar lattice bosons in the presence of random diagonal disorder},
  author = {Zhang, C. and Safavi-Naini, A. and Capogrosso-Sansone, B.},
  journal = {Phys. Rev. A},
  volume = {97},
  issue = {1},
  pages = {013615},
  numpages = {5},
  year = {2018},
  month = {Jan},
  publisher = {American Physical Society},
  doi = {10.1103/PhysRevA.97.013615},
  url = {https://link.aps.org/doi/10.1103/PhysRevA.97.013615}
}

@article{PhysRevA.107.043318,
  title = {Quantum phases of lattice dipolar bosons coupled to a high-finesse cavity},
  author = {Hebib, Yaghmorassene and Zhang, Chao and Yang, Jin and Capogrosso-Sansone, Barbara},
  journal = {Phys. Rev. A},
  volume = {107},
  issue = {4},
  pages = {043318},
  numpages = {7},
  year = {2023},
  month = {Apr},
  publisher = {American Physical Society},
  doi = {10.1103/PhysRevA.107.043318},
  url = {https://link.aps.org/doi/10.1103/PhysRevA.107.043318}
}

@article{PhysRevA.90.043635,
  title = {Quantum phases of dipolar soft-core bosons},
  author = {Grimmer, D. and Safavi-Naini, A. and Capogrosso-Sansone, B. and S\"oyler, \ifmmode \mbox{\c{S}}\else \c{S}\fi{}. G.},
  journal = {Phys. Rev. A},
  volume = {90},
  issue = {4},
  pages = {043635},
  numpages = {5},
  year = {2014},
  month = {Oct},
  publisher = {American Physical Society},
  doi = {10.1103/PhysRevA.90.043635},
  url = {https://link.aps.org/doi/10.1103/PhysRevA.90.043635}
}

@article{PhysRevA.90.043604,
  title = {Quantum phases of hard-core dipolar bosons in coupled one-dimensional optical lattices},
  author = {Safavi-Naini, A. and Capogrosso-Sansone, B. and Kuklov, A.},
  journal = {Phys. Rev. A},
  volume = {90},
  issue = {4},
  pages = {043604},
  numpages = {13},
  year = {2014},
  month = {Oct},
  publisher = {American Physical Society},
  doi = {10.1103/PhysRevA.90.043604},
  url = {https://link.aps.org/doi/10.1103/PhysRevA.90.043604}
}

@article{Capogrosso-Sansone:2011aa,
	author = {Capogrosso-Sansone, B. and Kuklov, A. B.},
	date = {2011/12/01},
	doi = {10.1007/s10909-011-0386-5},
	id = {Capogrosso-Sansone2011},
	isbn = {1573-7357},
	journal = {Journal of Low Temperature Physics},
	number = {5},
	pages = {213--226},
	title = {Superfluidity of Flexible Chains of Polar Molecules},
	url = {https://doi.org/10.1007/s10909-011-0386-5},
	volume = {165},
	year = {2011}}

@article{PhysRevA.103.043333,
  title = {Ground states of two-dimensional tilted dipolar bosons with density-induced hopping},
  author = {Zhang, Chao and Zhang, Jin and Yang, Jin and Capogrosso-Sansone, Barbara},
  journal = {Phys. Rev. A},
  volume = {103},
  issue = {4},
  pages = {043333},
  numpages = {8},
  year = {2021},
  month = {Apr},
  publisher = {American Physical Society},
  doi = {10.1103/PhysRevA.103.043333},
  url = {https://link.aps.org/doi/10.1103/PhysRevA.103.043333}
}

@article{Winding,
	Author = {Pollock, E L and Ceperley, D M},
	Doi = {10.1103/PhysRevB.36.8343},
	Journal = {Physical Review B},
	Language = {English},
	Month = dec,
	Number = {16},
	Pages = {8343--8352},
	Publisher = {American Physical Society},
	Title = {{Path-integral computation of superfluid densities}},
	Uri = {\url{papers3://publication/doi/10.1103/PhysRevB.36.8343}},
	Url = {http://journals.aps.org/prb/abstract/10.1103/PhysRevB.36.8343},
	Volume = {36},
	Year = {1987}}

@article{Kohn1959,
  author = {W. Kohn},
  title = {Analytic Properties of Bloch Waves and Wannier Functions},
  journal = {Phys. Rev.},
  volume = {115},
  pages = {809},
  year = {1959},
  month = aug
}

@article{zhang2022,
  title = {Supersolid phases of lattice dipoles tilted in three dimensions},
  author = {Zhang, Jin and Zhang, Chao and Yang, Jin and Capogrosso-Sansone, Barbara},
  journal = {Phys. Rev. A},
  volume = {105},
  issue = {6},
  pages = {063302},
  numpages = {9},
  year = {2022},
  month = {Jun},
  publisher = {American Physical Society},
  doi = {10.1103/PhysRevA.105.063302},
  url = {https://link.aps.org/doi/10.1103/PhysRevA.105.063302}
}

@article{PROKOFEV1998253,
title = {“Worm” algorithm in quantum Monte Carlo simulations},
journal = {Physics Letters A},
volume = {238},
number = {4},
pages = {253-257},
year = {1998},
issn = {0375-9601},
doi = {https://doi.org/10.1016/S0375-9601(97)00957-2},
url = {https://www.sciencedirect.com/science/article/pii/S0375960197009572},
author = {N.V Prokof'ev and B.V Svistunov and I.S Tupitsyn}
}

@article{Griesmaier2005,
  author = {A. Griesmaier and J. Werner and S. Hensler and J. Stuhler and T. Pfau},
  title = {Bose-Einstein Condensation of Chromium},
  journal = {Phys. Rev. Lett.},
  volume = {94},
  pages = {160401},
  year = {2005},
  doi = {10.1103/PhysRevLett.94.160401}
}

@article{Frisch2015,
  author = {A. Frisch and M. Mark and K. Aikawa and S. Baier and R. Grimm and A. Petrov and S. Kotochigova and G. Quéméner and M. Lepers and O. Dulieu and F. Ferlaino},
  title = {Ultracold Dipolar Molecules Composed of Strongly Magnetic Atoms},
  journal = {Phys. Rev. Lett.},
  volume = {115},
  pages = {203201},
  year = {2015},
  doi = {10.1103/PhysRevLett.115.203201}
}

@article{Su2023,
  author = {L. Su and A. Douglas and M. Szurek and R. Groth and S. F. Ozturk and A. Krahn and A. H. Hébert and G. A. Phelps and S. Ebadi and S. Dickerson and F. Ferlaino and O. Marković and M. Greiner},
  title = {Dipolar quantum solids emerging in a Hubbard quantum simulator},
  journal = {Nature},
  volume = {622},
  pages = {724--729},
  year = {2023},
  doi = {10.1038/s41586-023-06614-3}
}

@article{Bhongale2013,
  author = {S. G. Bhongale and L. Mathey and E. Zhao and S. F. Yelin and M. Lemeshko},
  title = {Quantum Phases of Quadrupolar Fermi Gases in Optical Lattices},
  journal = {Phys. Rev. Lett.},
  volume = {110},
  number = {15},
  pages = {155301},
  year = {2013},
  month = apr,
  doi = {10.1103/PhysRevLett.110.155301}
}

@article{Brunn2014,
  author = {J. K. Block and G. M. Bruun},
  title = {Properties of the density-wave phase of a two-dimensional dipolar Fermi gas},
  journal = {Phys. Rev. B},
  volume = {90},
  number = {15},
  pages = {155102},
  year = {2014},
  month = oct,
  doi = {10.1103/PhysRevB.90.155102}
}

@article{Yamaguchi2010,
  author = {Y. Yamaguchi and T. Sogo and T. Ito and T. Miyakawa},
  title = {Density-wave instability in a two-dimensional dipolar Fermi gas},
  journal = {Phys. Rev. A},
  volume = {82},
  number = {1},
  pages = {013643},
  year = {2010},
  month = jul,
  doi = {10.1103/PhysRevA.82.013643}
}

@article{Koziol2024,
  author = {K. Koziol and G. Morigi and R. Schmidt},
  title = {Quantum phases of hardcore bosons with repulsive dipolar interactions on two-dimensional lattices},
  journal = {SciPost Phys.},
  volume = {17},
  number = {4},
  pages = {111},
  year = {2024},
  doi = {10.21468/SciPostPhys.17.4.111}
}

@article{Baranov2008,
  author = {M. A. Baranov},
  title = {Theoretical progress in many-body physics with ultracold dipolar gases},
  journal = {Phys. Rep.},
  volume = {464},
  pages = {71--111},
  year = {2008},
  month = aug,
  doi = {10.1016/j.physrep.2008.04.007}
}

@article{Gadway2016,
doi = {10.1088/0953-4075/49/15/152002},
url = {https://doi.org/10.1088/0953-4075/49/15/152002},
year = {2016},
month = {jun},
publisher = {IOP Publishing},
volume = {49},
number = {15},
pages = {152002},
author = {Gadway, Bryce and Yan, Bo},
title = {Strongly interacting ultracold polar molecules},
journal = {Journal of Physics B: Atomic, Molecular and Optical Physics}
}

@article{Lu2009,
  author = {Mingwu Lu and Seo Ho Youn and Benjamin L. Lev},
  title = {Trapping ultracold dysprosium: a highly magnetic gas for dipolar physics},
  journal = {Phys. Rev. Lett.},
  volume = {104},
  pages = {063001},
  year = {2010},
  month = feb,
  doi = {10.1103/PhysRevLett.104.063001}
}

\end{document}